# TUNNELING AND RESONANT CONDUCTANCE IN ONE-DIMENSIONAL MOLECULAR STRUCTURES

M.A.Kozhushner, and V.S.Posvyanskii
Institute of Chemical Physics RAS, Moscow, Russia

I.I. Oleynik,
Department of Physics, University of South Florida, Tampa FL, USA

**Abstract**

We present a theory of tunneling and resonant transitions in one-dimensional molecular systems which is based on Green's function theory of electron sub-barrier scattering off the structural units (or functional groups) of a molecular chain. We show that the many-electron effects are of paramount importance in electron transport and they are effectively treated using a formalism of sub-barrier scattering operators. The method which calculates the total scattering amplitude of the bridge molecule not only predicts the enhancement of the amplitude of tunneling transitions in course of tunneling electron transfer through one-dimensional molecular structures but also allows us to interpret conductance mechanisms by calculating the bound energy spectrum of the tunneling electron, the energies being obtained as poles of the total scattering amplitude of the bridge molecule. We found that the resonant tunneling via bound states of the tunneling electron is the major mechanism of electron conductivity in relatively long organic molecules. The sub-barrier scattering technique naturally includes a description of tunneling in applied electric fields which allows us to calculate I-V curves at finite bias. The developed theory is applied to explain experimental findings such as bridge effect due to tunneling through organic molecules, and threshold versus Ohmic behavior of the conductance due to resonant electron transfer.

## 1. Introduction

In the recent years substantial research efforts were directed towards understanding the mechanisms of electron transport in one-dimensional structures such as metallic nanowires, carbon nanotubes, and single organic molecules [1,2]. This interest is motivated by active experimental investigations of one-dimensional nanostructures as perspective electronic device elements that would allow the fundamental scaling limitations of silicon-based electronics to be overcome [3]. Molecular electronics based on organic polymers, oligomers, and small organic molecules exploits intriguing electric properties of single molecules with the aim to reach the ultimate limit of miniaturization by producing single molecule diodes, transistors, and switches [4-7]. In order to fully utilize the unique properties of these one-dimensional molecular nanostructures, a fundamental understanding of conduction mechanisms in such systems has to be achieved.

The transport properties of single molecular devices are remarkably different from the electrical behavior of traditional solid-state devices. In general, most organic materials including single molecules do not conduct electricity in the usual way as it occurs in metals, i.e. by moving free electrons at applied bias (band conductivity). Due to the discrete nature of the electron energy spectrum and the presence of a substantial gap between occupied and unoccupied molecular energy levels, transport of electrons in the molecule attached to the metallic electrodes usually occurs via electron tunneling.

In many experiments, strong amplification of the tunneling current through organic molecules was observed. Without a bridging molecule, vacuum tunneling between metallic electrodes would exhibit exponential dependence of the tunneling current as a function of the distance between the electrodes, the tunneling exponent $2\left(2m_e W/\hbar^2\right)^{1/2} \sim 2.2 \text{ Å}^{-1}$ being determined by the work function of the electrodes $W \approx 4-5 \text{ eV}$. When a molecule is inserted between the electrodes, tunneling current dependence on the length of the bridging molecule exhibits exponential dependence with a much smaller tunneling exponent $\sim (0.4-1.2) \text{ Å}^{-1}$ [8]. This phenomenon is known in chemistry as the bridge effect and is primarily responsible for electron transfer reactions in biological systems [9] as well as electron tunneling through self-assembled monolayers in STM experiments [8,10].

Another important experimental discovery was made by several groups who observed extremely high values of the electric conductance through very long (tens and hundreds of nanometers) segments of DNA [12-15] with week dependence on the segment length. DNA is the organic molecule with a large HOMO-LUMO gap ($\sim 7-9 \text{ eV}$). Even taking into account the bridge effect, the tunneling current through the molecule would be practically zero already at lengths of several nanometers. But typical values of the conductance measured in many experiments are comparable with that of a metallic wire of the same thickness and length.

More surprisingly, different experiments demonstrated differing characteristics of I-V dependence. In particular, the I-V curve in experiment on bundles of DNA strands [12,14,15] showed an Ohmic behavior, i.e. the conductance was approximately constant for small biases.

However, in another experiment [13], the electrical transport through a 10 nm long single double helix DNA segment had shown a pronounced threshold as a function of applied bias. Other groups lately reproduced both the Ohmic and threshold I-V features of DNA conductance. In addition, the threshold in I-V curves was observed in other organic systems such as recently discovered di-block oligomers [16,17] that showed diode type I-V behavior.

The wide spectrum of the electrical behavior induced speculations in the scientific community about either the "insulating", "semiconducting", or "metallic" character of electron transport in organic molecules. However, it is important to make a distinction between electron transport in traditional bulk solid-state materials such as semiconductors or metals and a pure organic medium (we exclude the case of heavily doped conducting organic polymers). These two types of materials are fundamentally different because organic molecules are insulating in a sense that they do not possess free carriers as opposed to the case of semiconductors and metals where electrical conductance is due to the movement of free carriers such as electrons and holes. Obviously, the absence of free carriers in an organic medium does not allow direct transfer of transport mechanisms operational in bulk inorganic materials to explain the electrical conductance in single organic molecules and requires development of new theoretical concepts to explain the fundamental mechanisms of electron transport in molecular systems.

In this paper we present a theory of tunneling and resonant transitions that is able to rationalize and explain the fundamental features of electron transport in molecular systems within a conceptually simple and unified framework. Our approach is based on Green's function theory of electron sub-barrier scattering [18-20] off the structural units (or functional groups) of a molecular chain. The concept of sub-barrier scattering allows us to treat effectively the many-body effects in electron tunneling through organic molecules using a formalism of sub-barrier scattering amplitudes [18-20]. The method which calculates the total scattering amplitude of the bridge molecule not only predicts the enhancement of the amplitude of tunneling transitions in course of tunneling electron transfer through one-dimensional molecular structures but also allows us to interpret conductance mechanisms by calculating the bound energy spectrum of the tunneling electron, the energies being obtained as poles of the total scattering amplitude of the bridge molecule. In particular, we found that the resonant tunneling via bound states of the tunneling electron is the major mechanism of electron conductivity in relatively long organic molecules.

We show that the many-electron effects are of paramount importance in electron transport and its inclusion is critical for attaining a quantitative description of electron transport in one-dimensional molecular nanostructures. The sub-barrier scattering technique naturally includes a description of tunneling in applied electric fields which allows us to calculate I-V curves at finite bias. The developed theory is applied to explain experimental findings such as bridge effect due to tunneling through organic molecules, threshold versus Ohmic behavior of the conductance due to resonant electron transfer, and temperature effects in electron transport through organic molecules.

## 2. Amplitude of the tunneling transition

Almost all observable electron tunneling transitions are the transitions between the states of a continuous spectrum. These are the continuum of electronic states in metallic electrodes in the case of a molecule attached to metallic electrodes or the quasi-continuous vibration spectra of donor and acceptor in the case of donor-acceptor tunneling transitions. In most of these cases the Fermi "golden rule" is an excellent approximation for calculation of the probability $W_{\mathbf{k}_l \mathbf{k}_r}(V)$ of the electron tunneling transition from the state $\mathbf{k}_l$ of the left electrode to the state $\mathbf{k}_r$ of the right electrode under applied bias $V$

$$W_{\mathbf{k}_l \mathbf{k}_r}(V) = 2\pi \left| A_{\mathbf{k}_l \mathbf{k}_r} \right|^2 \delta\left( E(\mathbf{k}_l) - E(\mathbf{k}_r) \right) \tag{1}$$

where $A_{\mathbf{k}_l \mathbf{k}_r}$ is the amplitude of the transition. In this paper we use the atomic system of units $\hbar = e = m_e = 1$. The Fermi "golden rule" works because the transition amplitude $A_{\mathbf{k}_l \mathbf{k}_r}$ is much smaller than the characteristic energy scale of the substantial change of the density of states of metallic electrodes.

It is important to understand that the tunneling transitions through organic molecules are essentially quasi-equilibrium phenomena. Three main mechanisms might result in a non-equilibrium situation: 1) non-equilibrium kinetic phenomena in left and right electron subsystems due to removal of a tunneling electron from the left electrode and its addition to the right electrode in the course of tunneling; 2) inelastic interactions of tunneling electrons with vibrational degrees of freedom of the molecule and 3) non-equilibrium occupations of electronic levels inside the molecule in the course of tunneling. All three factors are of minor importance in the case of relatively small ($\sim 10^{-10} - 10^{-8}$ A) currents flowing through the molecule. For example, the time interval between individual electron tunneling transitions $\tau \sim 10^{-11} - 10^{-9}$ s is much larger that the characteristic time for establishment of thermal equilibrium in the electron gas of the metallic electrodes $\tau_e \sim 10^{-16} - 10^{-15}$ s and we can consider the electronic subsystem to be unperturbed by occasional electron tunneling transitions from the left to the right electrodes. In addition, the probability of vibrational excitations of the molecule in the course of tunneling transitions is small, therefore, we consider tunneling transitions as purely elastic. It is also worth mentioning that highly sophisticated non-equilibrium Green's function techniques widely used to describe transport in molecules [21,22] is an overcomplication of essentially quasi-equilibrium tunneling phenomena in one-dimensional molecular nanostructures.

Using equilibrium statistics of electrons in the right and left electrodes we can write an expression for the tunneling current through the molecule as

$$I(V) = 2\pi \int d\mathbf{k}_l d\mathbf{k}_r \left| A_{\mathbf{k}_l \mathbf{k}_r} \right|^2 \left[ f\left(\varepsilon(\mathbf{k}_l), T\right) - f\left(\varepsilon(\mathbf{k}_r) - V, T\right) \right], \qquad (2)$$

where $f\left(\varepsilon(\mathbf{k}_l), T\right)$ and $f\left(\varepsilon(\mathbf{k}_r) - V, T\right)$ are the Fermi distribution functions for electrons in the left and right electrodes, and $V$ is applied bias. The tunneling amplitude $A_{\mathbf{k}_l \mathbf{k}_r}$ is expressed via the one-electron Green's function of the molecule $G(\mathbf{r}', \mathbf{r}; \varepsilon)$ and the electron wave functions $\psi_{r,l}$ and electron potentials $U_{r,l}$ of right and left electrodes respectively:

$$A_{\mathbf{k}_l \mathbf{k}_r} = \int d\mathbf{r} \int d\mathbf{r}' \psi_l(\mathbf{r}', \mathbf{k}_l) U_l(\mathbf{r}') G(\mathbf{r}', \mathbf{r}; \varepsilon) U_r(\mathbf{r}) \psi_r(\mathbf{r}, \mathbf{k}_r), \qquad (3)$$

where $G(\mathbf{r}', \mathbf{r}; \varepsilon)$ also includes the electric field applied between the electrodes. It is also worth noting that the Fermi golden rule (1)-(2) combined with expression (3) for the amplitude of electron transition is formally equivalent to Bardeen's transfer Hamiltonian theory[[23] which is widely used to describe tunneling phenomena in condensed matter systems.

Our approach for calculating the current through the molecule is similar to the Landauer-Buttiker formalism or its generalization for the case of finite biases, non-equilibrium Green's function (NEGF) technique [24]. In both cases the current is expressed via a single-electron Green's function of a molecular system and the self-energies (or interface potentials $U_{r,l}$ as in our case) that take into account electronic interactions of the molecule with the electrodes, see (2),(3). However, in contrast to standard NEGF theory where the Green's function of a neutral molecule is evaluated, we work with the Green's function of a molecule plus an extra electron (or negative ion) which corresponds to the physical situation of an extra electron interacting with the molecule during the course of an electron transition from one electrode to another. At negative electron energies close to the Fermi energies of the electrodes, the Green's function of the negative molecular ion is mostly determined by the electronic states of a continuous spectrum, i.e. by the states of the electron scattered off the molecule. Therefore, the developed theory of sub-barrier scattering takes this contribution into account naturally within its remit by expressing Green's function via operators of sub-barrier scattering. In standard NEGF approaches based on the inversion of the Hamiltonian matrix of a neutral molecule, the continuous spectrum is completely ignored.

Let us first consider the case of small applied biases. In the case of tunneling through vacuum (no molecule present between electrodes) the one-electron Green's function is

$$G_0(\mathbf{r},\mathbf{r}';\varepsilon) = -\frac{1}{2\pi|\mathbf{r}-\mathbf{r}'|}\exp(-\kappa|\mathbf{r}-\mathbf{r}'|), \qquad (4)$$

where $\kappa = (2|\varepsilon|)^{1/2}$ is the tunneling exponent in vacuum (the energies of the tunneling electron are negative in respect to the vacuum energy level which is chosen as zero energy). The Green's function of a single molecular structural unit placed between electrodes is [18-20]

$$G(\mathbf{r},\mathbf{r}';\ \varepsilon) = G_0(\mathbf{r},\mathbf{r}';\ \varepsilon) + G_0(\mathbf{r},\mathbf{R}_1;\ \varepsilon)(-2\pi a(\varepsilon,\vartheta))G_0(\mathbf{R}_1,\mathbf{r}';\ \varepsilon) \qquad (5)$$

where $a(\varepsilon,\vartheta)$ is the electron sub-barrier scattering amplitude off this structural unit or center located at $\mathbf{R}_1$ and $\vartheta$ is the scattering angle between vectors $\mathbf{R}_1-\mathbf{r}$ and $\mathbf{r}'-\mathbf{R}_1$. The expression (5) has a clear physical meaning: the electron tunnels free plus scatters off the center located at $\mathbf{R}_1$. The amplitude of potential scattering can be obtained as an analytic continuation of the amplitude at positive electron energies $\varepsilon > 0$ or real momenta to the region of negative energies or imaginary momenta $k = i\kappa = i\sqrt{2|\varepsilon|}$.

We have developed a variational asymptotic method for calculation of tunneling scattering amplitudes [18-20]. Within this approach the exponential tail of the wave function of the system of $n_c +1$ electrons ($n_c$ electrons of the center plus one tunneling electron) is determined by varying the total energy functional. Therefore, the exchange interaction between the tunneling electron and the electrons of the scattering center is explicitly taken into account. The general form of sub-barrier scattering amplitude $a(\varepsilon,\vartheta)$ is [25,20]:

$$a(\varepsilon,\ \vartheta) = \frac{a_{pole}(\vartheta)}{\varepsilon - \varepsilon_0} + a_{pot}(\varepsilon,\ \vartheta) \qquad (6)$$

where the energy $\varepsilon_0$ in the pole term is the energy of the virtual bound state of the tunneling electron which is formed in course of scattering off the center. The second term in (6) is the potential part of the scattering amplitude which is an analytic and smooth function of energy. The energy and angular dependencies of the sub-barrier scattering amplitude off the atoms $H$ (singlet and triplet scattering), $He$, $Ne$, $Be$ [20] are shown in Fig. 1. The hydrogen and beryllium atoms are the examples of open shell systems that exhibit strong scattering, whereas closed shell systems such as helium and neon exhibit an order of magnitude weaker scattering. In all cases the scattering amplitude $a(\varepsilon,\vartheta)$ shows similar energy and angular behavior. The potential part $a_{pot}(\varepsilon,\vartheta)$ exhibits a rapid monotonic increase for systems with effective repulsion when $a_{pot}(\varepsilon,\vartheta) > 0$ (triplet H, Be, Ne, He) and a monotonic decrease for systems with an effective attraction when $a_{pot}(\varepsilon,\vartheta) < 0$ (singlet scattering off the hydrogen atom). For example, in case of triplet electron scattering off the hydrogen atom, $a_{pot}(\varepsilon,\vartheta)$ increases from $-30$ to $-2.4$ when $\varepsilon$ varies from $0.18\,\text{a.u.} = -5\,\text{eV}$ to $0$. The minimum and subsequent increase of the total scattering amplitude is a result of approaching the energy of the bound state $\varepsilon_0$ (see (6)) when the pole term starts to contribute to $a(\varepsilon,\vartheta)$. In addition, $a(\varepsilon,\vartheta)$ strongly depends on the scattering angle $\vartheta$ if the energy $\varepsilon$ is within the tunneling energy interval, but this dependence disappears when $|\varepsilon| \to 0$.

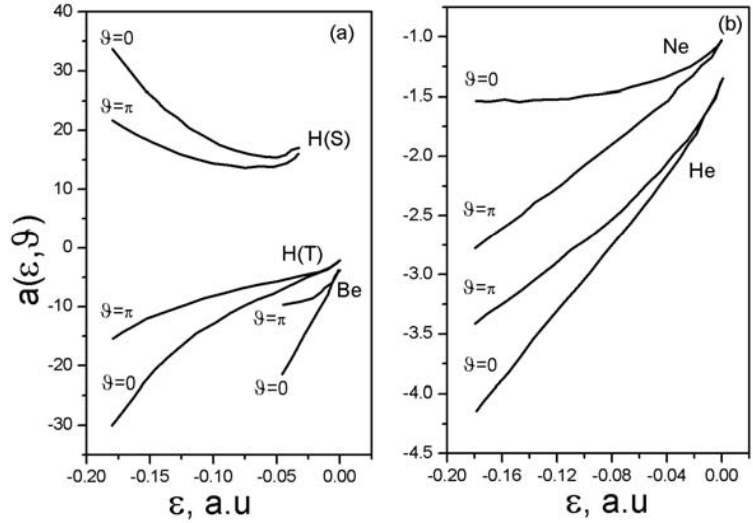

Fig. 1. The energy dependence of the amplitude of sub-barrier scattering $a(\varepsilon,\vartheta)$ at two scattering angles $\vartheta = 0$, and, $\vartheta = \pi$ for (a) singlet and triplet scattering off the hydrogen atom, beryllium, and (b) helium and neon.

The expression for a tunneling Green's function for the general case of a molecular chain consisting of several scattering centers is

$$G = G_0 + G_0 \hat{T} G_0, \tag{7}$$

where $\hat{T}$ is the total scattering operator which is related to the total scattering amplitude as $\hat{T} = -2\pi A$. The scattering operator depends on coordinates $\mathbf{r}, \mathbf{r}'$ via scattering angles for the electron coming from point $\mathbf{r}$, scattered off the centers $\{\mathbf{R}_1, ..., \mathbf{R}_N\}$ and arriving at point $\mathbf{r}'$. The total scattering operator $\hat{T}$ is determined via solution of the system of $N$ linear equations [20]

$$\hat{T}_n(\varepsilon) = \hat{t}_n(\varepsilon) + \sum_{\substack{k=1, \\ k \neq n}}^{N} \hat{t}_n(\varepsilon) G_0(\mathbf{R}_n, \mathbf{R}_k; \varepsilon) \hat{T}_k, \quad n = 1, ..., N \tag{8}$$

as

$$\hat{T}(\varepsilon) = \sum_{n=1}^{N} \hat{T}_n(\varepsilon), \tag{9}$$

where $\hat{t}_n \equiv -2\pi a_n(\varepsilon, \vartheta) \delta(\mathbf{R} - \mathbf{R}_n)$ is the scattering operator off the individual $n$th center, and the $n$th component $T_n$ of the solution vector $\{T_1, ..., T_N\}$ is the partial scattering operator that gives a subset of all multiple scattering events that start from center $n$. In deriving (8), we assumed that the distance between any two centers is sufficiently large so that the effective short-range potentials of the centers weekly overlap. The system (8) can be generalized for the case of non-zero overlap of the centers.

The tunneling Green's function of the system of $N$ centers contains all the possible multiple scattering events

$$G(\mathbf{r}, \mathbf{r}'; \varepsilon) = G_0(\mathbf{r}, \mathbf{r}'; \varepsilon) + \sum_{i,k} G_0(\mathbf{r}, \mathbf{R}_i; \varepsilon) \Gamma(\mathbf{R}_i, \mathbf{R}_k) G_0(\mathbf{R}_k, \mathbf{r}'; \varepsilon), \tag{10}$$

$$\Gamma(\mathbf{R}_i, \mathbf{R}_k; \varepsilon) = \begin{cases} \dfrac{1}{D} \{t_i G_0(\mathbf{R}_i, \mathbf{R}_\alpha; \varepsilon) t_\alpha G_0(\mathbf{R}_\alpha, \mathbf{R}_\beta) t_\beta .... t_\nu G_0(\mathbf{R}_\nu, \mathbf{R}_k; \varepsilon) t_k\}, & i \neq k \\ \dfrac{1}{D} t_i, & i = k \end{cases} \tag{11}$$

where $D$ is the determinant of the system (8), and greek indices $\alpha, \beta, ..., \nu$ enumerate all the intermediate centers other than endpoints $i$ and $k$. Each term in the sum (10) can be easily visualized using graphical diagrams: each graph connects space points $\mathbf{r}$ and $\mathbf{r}'$ by all the possible paths running through the centers $\{\mathbf{R}_1, \mathbf{R}_2, ..., \mathbf{R}_N\}$. The vertex of each graph at the center $\mathbf{R}_i$ is represented by the center's scattering operator $t_i(\varepsilon, \vartheta_i)$ which depends on the scattering angle $\vartheta_i$, and the segment connecting any two centers centers $i$ and $k$ is represented by the vacuum Green's function $G_0(|\mathbf{R}_i - \mathbf{R}_k|; \varepsilon)$. In Fig. 2 we show all the diagrams contributing to the tunneling Green's function of the system consisting of 2 scattering centers.

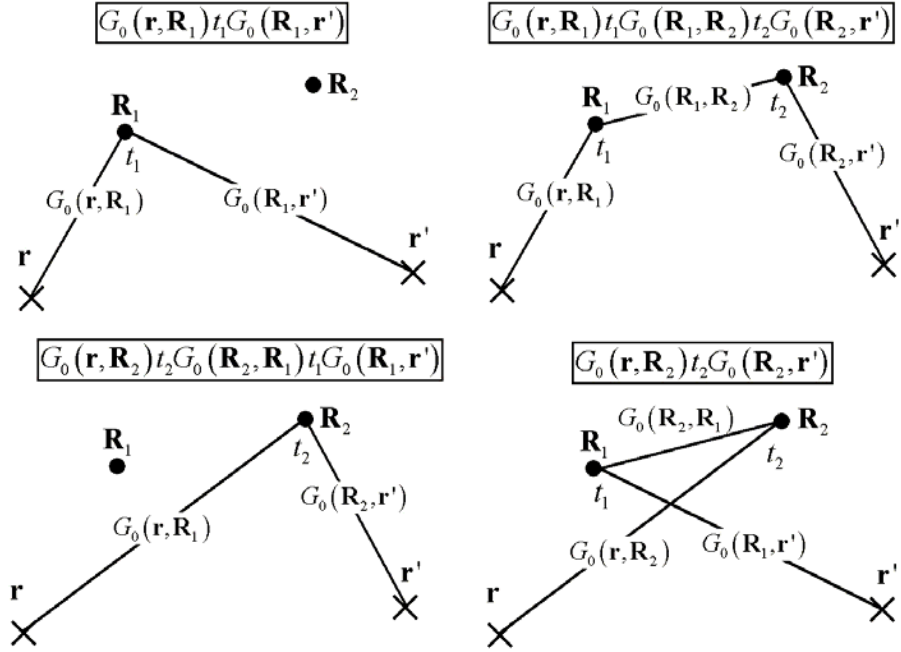

Fig. 2. Graphs contributing to the tunneling Green's function for a system of two centers. Top panel corresponds to partial scattering operator $T_1$, bottom panel – to $T_2$.

The poles of the total scattering operator constitute the spectrum of the bound states of the tunneling electron. They are easily calculated as roots of the determinant $D$ of the system (8). The graphs contributing to $D$ are all possible self-returning paths starting from each scattering center. Based on a knowledge of the energy spectrum, specifically, the position of the energy spectrum in respect of the Fermi energies of the left and right electrodes, we can identify different mechanisms of transport through the molecule such as ordinary tunneling and resonant electron transfer mechanisms.

## 3. Ordinary tunneling

When the bound energy spectrum of the tunneling electrons is higher than the Fermi energies of both the left and right electrodes, the ordinary tunneling is the major mechanism of electron transport through the molecule. This regime is characterized by the exponential dependence of the tunneling current along the length of the molecule, $I \propto \exp(-2\kappa L)$. The physical consequence of the interaction of the tunneling electron with the molecule is the substantial reduction of the tunneling exponent $\kappa$ compared to that for tunneling in vacuum $\kappa_0 = \sqrt{2|\varepsilon_F|} \simeq \sqrt{2W} \sim 1.1 \, \text{Å}^{-1}$, where $W \sim 4-5 \, \text{eV}$ is the work function of the metallic electrodes.

The total tunneling amplitude of this transition is obtained by substituting the expression for the tunneling Green's function (10) in the expression for the tunneling amplitude (3). It is possible to show that the general expression for tunneling amplitude $A_{lr}$ can be written as:

$$A_{lr} = \widetilde{A_{lr}} G_0\left(|\mathbf{R}_l - \mathbf{R}_r|;\varepsilon\right)\left\{1 + B\left(|\mathbf{R}_l - \mathbf{R}_r|;\varepsilon\right)\right\} \tag{12}$$

where the prefactor $\widetilde{A_{lr}}$ is

$$\widetilde{A_{lr}}(\varepsilon,\mathbf{k}_l,\mathbf{k}_r) = \left\{\int d\mathbf{r}\, \psi_l(\mathbf{r},\mathbf{k}_l) U_l(\mathbf{r}) G_0\left(|\mathbf{R}_l - \mathbf{r}|;\varepsilon\right)\right\}\left\{\int d\mathbf{r}'\, \psi_r(\mathbf{r}',\mathbf{k}_r) U_r(\mathbf{r}) G_0\left(|\mathbf{R}_r - \mathbf{r}|;\varepsilon\right)\right\} \tag{13}$$

and the bridge enhancement factor has exponential dependence on the length of the molecule

$$B\left(|\mathbf{R}_l - \mathbf{R}_r|;\varepsilon\right) = \exp\left(+\beta(\varepsilon)|\mathbf{R}_l - \mathbf{R}_r|\right). \tag{14}$$

The expression (14) was derived for the molecules that consist of more than five centers, and is not valid for very short molecules.

The dependence of the bridge enhancement exponent $\beta(\varepsilon)$ on $\varepsilon$ for a model system consisting of the chain of hydrogen atoms separated by the distance $d = 6$ a.u. for the case of singlet scattering $a_s(\varepsilon,\vartheta) > 0$ is shown in Fig. 3. For this particular case, the lowest energy level of the bound spectrum is close to vacuum level (zero energy). Therefore, the pole term of the scattering amplitude of each individual center (6) gives a small contribution at $\varepsilon \sim \varepsilon_F$ and the physics of the bridge effect (amplification of the tunneling current) is mainly determined by the potential term $a_{pot}(\varepsilon, \vartheta)$. In order to demonstrate this, the bridge enhancement exponent was calculated for the case when only the pole term $a_{pole}(\vartheta)/(\varepsilon - \varepsilon_0)$ was included and as is seen from Fig. 3 (dotted curve), the pole contribution is very small. However, its contribution increases as the tunneling energy approaches to the lowest energy level of the bound spectrum, $\varepsilon \approx -0.025$ a.u., see Fig. 3.

The qualitative behavior of the bridge enhancement exponent can be understood by considering the specific case that allows us to obtain an analytic solution for $B\left(|\mathbf{R}_l - \mathbf{R}_r|;\varepsilon\right)$.

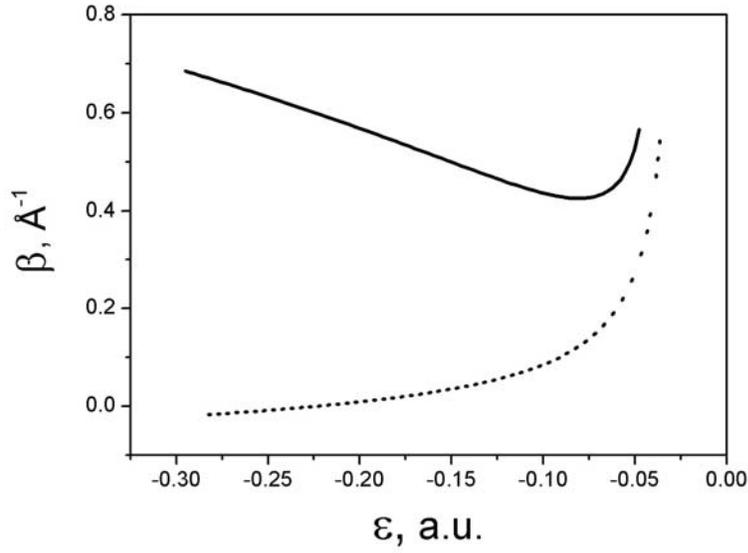

Fig. 3. Exponent $\beta$ of the bridge enhancement factor as a function of electron tunneling energy (bold curve). Dotted curve is for the case when only the pole term $a_{pole}(\vartheta)/(\varepsilon - \varepsilon_1)$ is taken into account.

In particular, if the energy of the tunneling electron is much lower than the bound energy spectrum and the scattering amplitude of individual centers is positive (effective attraction), then the following condition is satisfied

$$a(\varepsilon,\vartheta)G_0(\mathbf{R}_i,\mathbf{R}_{i+1};\varepsilon) = \frac{a(\varepsilon,\vartheta)}{d}\exp(-\kappa d) \ll 1 \qquad (15)$$

and the bridge enhancement factor is

$$B(\mathbf{R}_{lr};\varepsilon) = C\left(1 + \frac{a(\varepsilon,\vartheta)}{d}\right)^{N+1}, \qquad (16)$$

where $N$ is the number of centers comprising the molecule, and the numerical coefficient $C \sim 1$. Within this model, the bridge enhancement exponent $\beta$ obtained from (16) is

$$\beta(\varepsilon) = \ln\left(1 + \frac{a(\varepsilon,\vartheta)}{d}\right). \qquad (17)$$

The expression (17) shows, that the energy dependence of the bridge enhancement exponent in Fig. 3 is determined entirely by the energy dependence of the scattering amplitude of the individual scattering center $a(\varepsilon, \vartheta)$. In particular, the minimum in $\beta(\varepsilon)$ shown in Fig. 3 is due to the minimum of singlet scattering amplitude $a(\varepsilon, \vartheta)$ of the hydrogen atom, see Fig. 1a.

The expression for the tunneling current in the case of a small applied bias $V \ll |\varepsilon_F| = W$ is obtained by substituting (12) and (14) in (2)

$$I(V) = V \left\{ \frac{\rho_l(\varepsilon_F) \rho_r(\varepsilon_F)}{2\pi} \widetilde{A}_{lr}^{\,2}(\varepsilon_F) \frac{\exp(-2\kappa_0 |\mathbf{R}_l - \mathbf{R}_r|)}{|\mathbf{R}_l - \mathbf{R}_r|^2} \left[ 1 + \exp(\beta(\varepsilon_F)|\mathbf{R}_l - \mathbf{R}_r|) \right]^2 \right\} \quad (18)$$

where $\varepsilon_F = -W$, $W$ is the work function of the left and right electrodes, $\rho_l(\varepsilon_F)$ and $\rho_r(\varepsilon_F)$ are their densities of states at the Fermi level, $\kappa_0 = \sqrt{2|\varepsilon_F|}$ and $\widetilde{A}_{lr}(\varepsilon_F)$ is the prefactor $\widetilde{A}_{lr}(\varepsilon, \mathbf{k}_l, \mathbf{k}_r)$ averaged over the energy surfaces $\varepsilon(\mathbf{k}_l) = \varepsilon(\mathbf{k}_r) = \varepsilon_F$. The expression in curly brackets is the tunneling conductance $G = I/V = dI/dV$ at $V = 0$.

## 4. Effect of an electric field

If a finite bias is applied, then the problem of sub-barrier scattering is solved in the presence of an electric field. We must address the issue of possible charge redistribution within the molecule due the applied electric field and a tunneling current passing through the molecule. An important reference is the magnitude of the microscopic electric field inside the molecule, $E_{mol} \sim \frac{e}{a_0^2} = 1\ a.u. \sim 10^{12}$ V/m. If several volts is applied across a molecule several nanometers long, then the magnitude of the external electric field is $1\text{V}/1\text{nm} = 10^9$ V/m which is at least three orders of magnitude smaller than the internal microscopic electric field. Therefore, the change of the scattering operator due to the modification of the local electronic structure by the external electric field is negligible. In addition, the external electric field due to an applied voltage is essentially an electrostatic electric field that would exist in a system of two bare electrodes without a molecule because (1) there are no external charges between electrodes present, (2) the molecular polarizability effects are of minor importance, and (3) tunneling current passing through the molecule is small. Then, in the case of flat electrodes and a linear molecule the external electrostatic potential is distributed linearly along the molecule and there is no need to solve self-consistently the Poisson equation. More over, the change of the total energy of the isolated molecule upon application of external electrostatic field is very small. We found that total energy of eight-ring thiophene oligomer changed only by 0.08 eV when the molecule was placed in the electric field corresponding to bias 2 V applied across the length of the molecule.

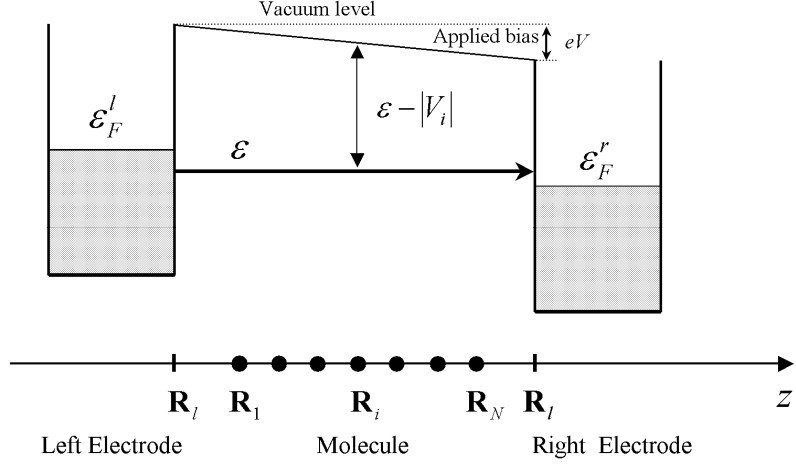

Fig. 4. Energy diagram for the left-electrode molecule right-electrode junction. Zero energy is the energy of the vacuum level.

Due to the above arguments, the effect of an electric field can be easily included into our formalism of sub-barrier scattering by parametric referencing of the local vacuum levels of the scattering centers by the local electrostatic potential and an additional modification of the vacuum Green's function to include explicitly the electrostatic potential. In particular, we assume that the polarity of applied bias is such that the electrons tunnel from the left electrode to the right electrode and the energy of $i$th scattering center is shifted as $\varepsilon_i \to \varepsilon_i + V_i$, where $V_i$ is the value of the local electrostatic potential at the center, $V_i = -V \frac{R_i}{R_{lr}} < 0$, see Fig. 4.

The vacuum Green's functions $G_0(\mathbf{R}_n, \mathbf{R}_k; \varepsilon)$ connecting individual scattering centers $R_n$ and $R_k$ in (8) are replaced by quasi-classical Green's functions for the electron in a homogeneous electric field $E = V/R_{lr}$

$$G_V(\mathbf{R}_n, \mathbf{R}_k; \varepsilon, E) = -\frac{1}{2\pi |\mathbf{R}_n - \mathbf{R}_k|} \cdot \exp(-S_V(\mathbf{R}_n, \mathbf{R}_k)), \quad (19)$$

where the action $S_V(\mathbf{R}_n, \mathbf{R}_k)$ is

$$S_V(\mathbf{R}_n, \mathbf{R}_k) = \int_{\mathbf{R}_k}^{\mathbf{R}_n} dz \sqrt{2(|\varepsilon| - Ez)} = \frac{2\sqrt{2}|\mathbf{R}_l - \mathbf{R}_r|}{3V} \left\{ (|\varepsilon| - |V_n|)^{3/2} - (|\varepsilon| - |V_k|)^{3/2} \right\} \quad (20)$$

Here we assumed that the scattering centers are along a straight line but it is easy to generalize the quasi-classical expression (19) to include a general 3-dimensional configuration of the molecule.

The system of linear equations for the total scattering operator is modified accordingly:

$$T_n(\varepsilon,V) = t_n(\varepsilon - V_n) + \sum_{k \neq n} t_n(\varepsilon - V_n) G_V(R_n, R_k; \varepsilon) T_k(\varepsilon,V), \tag{21}$$

where the total scattering operator is

$$T(\varepsilon,V) = \sum_{n=1}^{N} T_n(\varepsilon,V). \tag{22}$$

It is easy to modify the expressions (10), (11) for the total Green's function and the expression for the tunneling amplitude to include explicitly the electric field. Then, assuming that $|V| < |\varepsilon_F|$ $|V| < |E_F|$, we can derive the expression for the tunneling current at finite bias $V$:

$$I(V) = \int_{\varepsilon_F - V}^{\varepsilon_F} d\varepsilon \rho_l(\varepsilon) \rho_r(\varepsilon - V) \widetilde{A}_{lr}^2(\varepsilon) \frac{\exp(-2S_V(\varepsilon, |\mathbf{R}_l - \mathbf{R}_r|))}{2\pi |\mathbf{R}_l - \mathbf{R}_r|} \left[1 + \exp(\beta(\varepsilon,V)|\mathbf{R}_l - \mathbf{R}_r|)\right]^2. \tag{23}$$

The I-V curve for the same model system, a molecular wire of hydrogen atoms separated by a distance $d = 6$ a.u., is shown in Fig. 5. We can interpret general features by examining (23) in the case of small biases ($|V| \ll |\varepsilon_F|$). The differential conductance for the case of vacuum tunneling (two electrodes without molecule) is

$$\frac{d(\ln I)}{dV} = -\kappa |\mathbf{R}_l - \mathbf{R}_r| \frac{V}{2|\varepsilon_F|}, \tag{24}$$

that is the tunneling current increases exponentially with bias $V$. For the case of tunneling through the molecule (bridge enhanced current) the differential conductance is

$$\frac{d(\ln I_t)}{dV} \approx -\kappa |\mathbf{R}_l - \mathbf{R}_r| \frac{V}{2|\varepsilon_F|} - 2|\mathbf{R}_l - \mathbf{R}_r| \frac{d\beta(\varepsilon,V)}{d|V|}. \tag{25}$$

Because the energy dependence of the bridge enhancement exponent $\beta(\varepsilon)$ is determined entirely by the energy dependence of the scattering amplitude $a(\varepsilon,\vartheta)$ of an individual center, see (17), we can make the conclusion that the derivative $d\beta(\varepsilon,V)/d|V|$ is always negative based on the fact of the general monotonic behavior of the scattering amplitude for both cases of attractive $a(\varepsilon,\vartheta) > 0$ and repulsive $a(\varepsilon,\vartheta) < 0$ interactions, see Fig. 1.

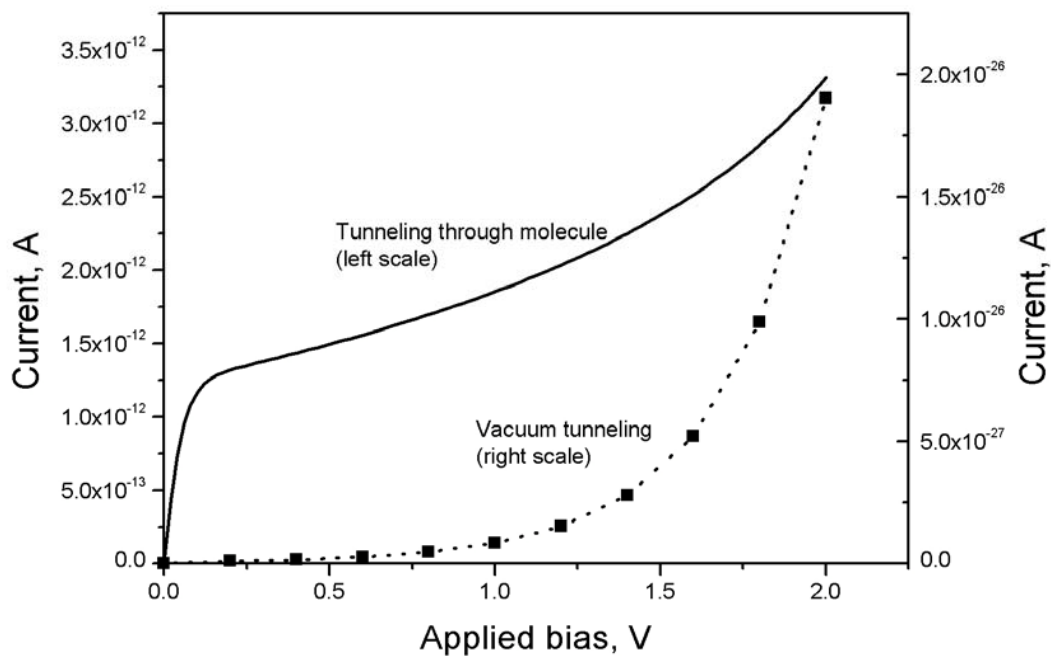

Fig. 5. Current-voltage curves for the case of vacuum tunneling (dotted line with squares, right scale) and for the tunneling through the molecule (left scale).

Therefore, the bridge enhanced current increases with bias much slower as compared to vacuum tunneling excluding the interval of very small bias, see Fig. 5. At the same time, the tunneling current through the molecule is enhanced by a factor $10^{15}$ (bridge enhancement) compared to vacuum tunneling.

The tunneling current in Fig. 5 was calculated for 20 $\overset{o}{A}$ long molecule and its value is pn the order of picoamperes. In most experiments, however, the measured tunneling current is on the order of nanoamperes for organic molecules of such length, which indicates the presence of another mechanism of electron transfer that substantially enhances the current through the molecule. This mechanism is the resonant electron transfer through the bound energy levels of the tunneling electron that we are going to consider in the next section.

## 5. Resonant electron transfer

If the bound energy spectrum of the tunneling electron is close to the Fermi energies of the electrodes, a new transport mechanism comes into play, see Fig. 6. The sub-barrier scattering formalism developed so far has to be modified to include resonant electron transfer. One of the critical pieces of information is the bound energy spectrum that is required to assess the states contributing to resonant electron transfer.

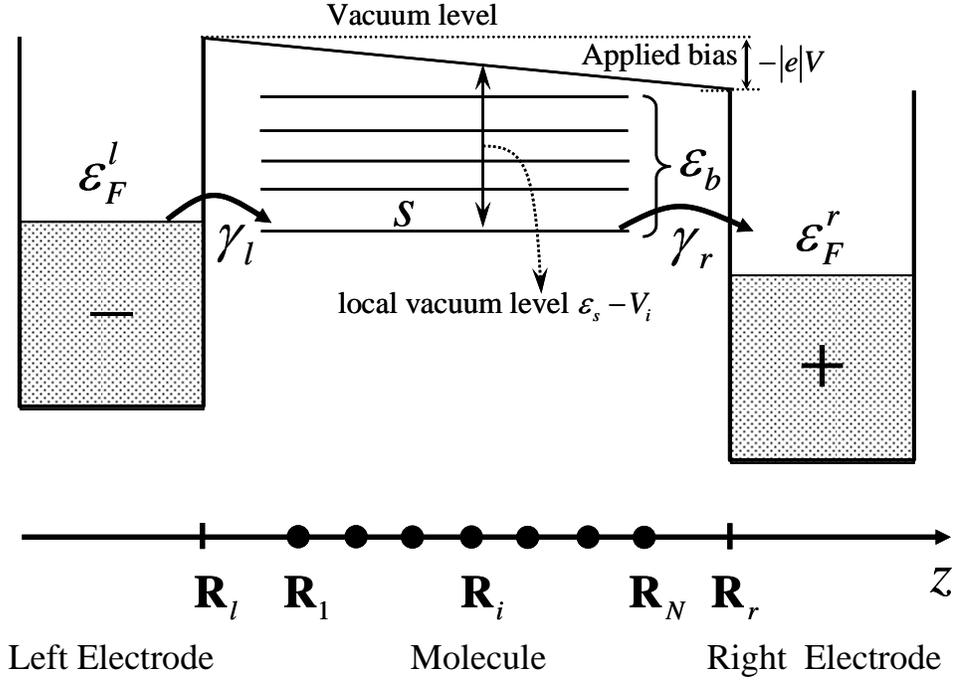

Fig. 6. Resonant tunneling transfer occurs when some energy levels of the bound spectrum are within energy interval $\varepsilon_r^F \leq \varepsilon_s \leq \varepsilon_l^F$, where $\varepsilon_r^F = \varepsilon_l^F - |V|$.

As was already mentioned, the bound energy spectrum is automatically obtained in course of solving system (21), which is solved in order to find the total scattering operator $T(\varepsilon, V)$. The energies of the bound states $\varepsilon_s$ in the electric field $E$ are found as poles of the total scattering operator, i.e. as the roots of the determinant of system (21)

$$\det|\delta_{ik} - t_i(\varepsilon - V_i, \vartheta)G_V(R_i, R_k; \varepsilon, V)| = 0. \tag{26}$$

The bound state wave-function $\psi_s$ corresponding to $s$-th root of the secular equation (26) is expressed via the normalized solution $\vec{T}(s) = \{T_1(s), T_2(s), ..., T_N(s)\}$ of homogeneous system (21) for $s$-th scattering vector

$$\psi_s(\mathbf{r}; \varepsilon_s) = \sum_l T_l(s) \tilde{\varphi}(\mathbf{r} - \mathbf{R}_l; \varepsilon_s), \tag{27}$$

where $\tilde{\varphi}(\mathbf{r} - \mathbf{R}_l)$ are additional contributions to the exponential tail of the electron wave function of the tunneling electron due to its interaction with scattering center $l$. These wave functions are determined in the course of a variational minimization procedure that was developed with a specific focus to resolve exponentially small contributions.

There is a clear analogy between functions $\tilde{\varphi}(\mathbf{r}-\mathbf{R}_l)$ that form the wave function of the tunneling electron (27) and the atomic wave functions that form the independent electron, tight-binding wave function (29) in LCAO method. The components $T_l(s)$ of the scattering vector $\vec{T}$ are similar to the coefficients $C_l$ of the LCAO expansion of the tight-binding wave functions. The LCAO energies and wave-functions for a one-dimensional chain are written as

$$\varepsilon_k = \varepsilon_0 - 2h\cos(kd) \tag{28}$$

$$\psi_k^{TB} = \sqrt{\frac{2}{N}} \sum_{l=1}^{N} C_l \varphi(\mathbf{r}-\mathbf{R}_l), \tag{29}$$

where the one-dimensional wave vector of a molecular wire is $k = \pi n/(N+1)d$, $n = 1,...,N$, $\varphi(\mathbf{r}-\mathbf{R}_l)$ is the atomic wave function centered at atom $l$ that has on-site energy $\varepsilon_0$, and $h$ is the hopping integral between nearest neighbor atoms. The LCAO expansion coefficients $C_l$ are determined via the solution of the tight-binding secular equation that includes explicitly the effect of the electric field by referencing the on-site energies of each center by the corresponding local electrostatic potential: $\varepsilon_0 \to \varepsilon_0 - |V_i|$. In the case of zero applied bias $C_l = \sin(kR_l)$, i.e. these are the usual Bloch wave phase factors for the wave function of the system with 1-d periodicity.

It is worth discussing the connection of many-electron sub-barrier scattering theory and the essentially one-electron tight-binding approach that gives the tight-binding energy spectrum (28) and the tight-binding wave functions (29). The tight-binding method is a simplified version of the standard density-functional LCAO method for an electronic structure widely used for the description of electronic transport in molecules. It is possible to show that if the distance between scattering centers is large, $d > 10$ a.u., the solution for the eigenspectrum and wave-functions of the tunneling electron obtained within the sub-barrier scattering approach can be cast into the tight-binding form (28) and (29), if the hopping integral $h$ is related to the parameters $a_{pole}$ and $\varepsilon_0$ of the pole term in (6) via expression

$$h = a_{pole}/d \cdot \exp\left(-\sqrt{2|\varepsilon_0|}d\right). \tag{30}$$

However, at smaller distances between the scattering centers the many electron effects substantially modify the physics of the resonant tunneling and the two approaches give drastically different results. In particular, within the tight-binding approach the probability of resonant tunneling is close to zero. In order to see this, it is necessary to write down the general form of the resonant tunneling amplitude $A_{res}(\varepsilon)$.

Let us consider one of the bound energy states $s$ that is in the resonance condition, i.e. $\varepsilon_F - |V| < \varepsilon_s < \varepsilon_F$, see Fig. 6. In contrast to the case of ordinary tunneling, we do not need to use the previous expression for the total Green's function (10) that explicitly takes into account the contribution of the entire energy spectrum of the tunneling electron including continuous states. Instead, only resonant state $\varepsilon_s$ makes a dominant contribution to the total Green's function which can be explicitly written down using spectral representation as

$$G_{res}(\mathbf{R},\mathbf{R}';\varepsilon) = \frac{\psi_s(\mathbf{R};\varepsilon_s)\psi_s(\mathbf{R}';\varepsilon_s)}{\varepsilon - \varepsilon_s + i\gamma_s}, \qquad (31)$$

where $\psi_s(\mathbf{R};\varepsilon_s)$ is the wave function of the bound state $s$ of the tunneling electron that is given by expression (27) and $\gamma_s$ is the imaginary part of the energy of resonant state $s$ associated with the finite lifetime of this energy level. Substituting (31) into the expression for the transition amplitude (3) we obtain the expression for the resonant amplitude $A_{res}(\varepsilon,\varepsilon_s)$

$$A_{res}(\varepsilon,\varepsilon_s) = \widetilde{A}_l(\varepsilon_s)\widetilde{A}_r(\varepsilon_s)\frac{T_1(s)T_N(s)}{\varepsilon - \varepsilon_s + i\gamma_s}, \qquad (32)$$

where amplitudes

$$\begin{aligned}\widetilde{A}_l(\varepsilon_s,\mathbf{k}_l) &= \int d\mathbf{r}\,\psi_l(\mathbf{r},\mathbf{k}_l)U_l(\mathbf{r})\widetilde{\varphi}(\mathbf{r}-\mathbf{R}_1;\varepsilon_s), \\ \widetilde{A}_r(\varepsilon_s,\mathbf{k}_r) &= \int d\mathbf{r}\,\psi_r(\mathbf{r},\mathbf{k}_r)U_r(\mathbf{r})\widetilde{\varphi}(\mathbf{r}-\mathbf{R}_N;\varepsilon_s),\end{aligned} \qquad (33)$$

describe the coupling of the electronic states $\mathbf{k}_l$ and $\mathbf{k}_r$ of the left and right electrodes with the resonant wave function $\widetilde{\varphi}(\mathbf{r}-\mathbf{R}_1;\varepsilon_s)$ of the tunneling electron. Obviously, the amplitude of the resonant tunneling transition (32) depends on partial scattering operators $T_1(s)$ and $T_N$ off the first and last centers of the molecular wire that are obtained as components of the solution of the system (21) in the case of sub-barrier scattering or as the first $C_1$ and last $C_N$ LCAO coefficients of the tight-binding solution (29). We would like to examine the spatial behavior of both partial scattering operators $T_l(s)$ and the LCAO coefficients $C_l(s)$ in order to see the difference in the two approaches in describing resonant tunneling.

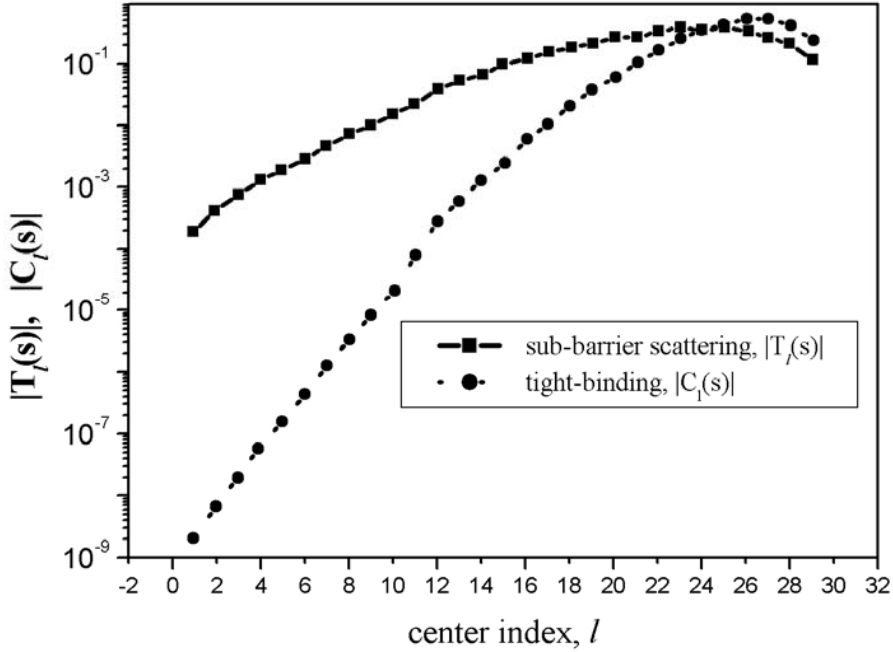

Fig. 7. Sub-barrier scattering vector $\vec{T}(s) = \{T_1, T_2, ..., T_N\}$ and tight-binding LCAO coefficients $C_l(s)$ along molecular wire.

In general, the tight-binding amplitude of resonant tunneling $A_{res}$ is negligible because the LCAO coefficient $C_1$ closest to the left electrode is exponentially small compared to the coefficient $C_N$ at the right electrode. In order to demonstrate this let us consider the specific example of a molecular chain of $N = 30$ centers each having on-site energy $\varepsilon_0 = 1$ eV and hopping integral $h = 1$ eV, the values being chosen to capture characteristic valence electronic interactions in organic molecules. We also assume that the Fermi energy of metallic electrodes at $V = 0$ is $\varepsilon_F = -4$ eV which corresponds to a typical work function of metals. Under these conditions the lowest energy level of the tight-binding band corresponding to the state with $k_1 = \pi/(N+1)d$ is 1 eV above the Fermi energy at zero bias. The resonant condition within the tight-binding model is satisfied if the applied bias is equal to the threshold value $V_{th} = 1.25$ eV, i.e. when the lowest energy level of the tight-binding band aligns with the Fermi level. The tight-binding LCAO coefficients $C_l(s)$ as a function of center index $l$ are shown in Fig. 7 (curve with dots). As is seen from Fig. 7, the tight-binding solution decays extremely fast when going from the right to the left electrodes, i.e. $C_1/C_{30} \approx 10^{-8}$. As we will see later, the amplitude of the resonant transition is proportional to the coefficient of the wave function at the first center close to the left electrode, i.e. $T_1$ in the case of sub-barrier scattering or $C_1$ in the case of tight-binding. Therefore, even if the resonant condition is attained in the course of a bias increase, the amplitude of the transition is negligible.

This numerical result has a simple physical explanation that is valid for any electronic structure method that uses a single electron approximation in describing electron-electron interactions within a potential framework, including DFT. The external electron passing through the molecule is essentially a weakly interacting, free-electron like particle. When the bias is applied, the electron in the lowest bound state with $\varepsilon \approx \varepsilon_F$ becomes localized near the right electrode and penetrates to the left under a triangular potential barrier $U = -Vz/R_{lr}$ with the effective mass $m^* = \hbar^2/(d^2\varepsilon/dk^2) = 1/2hd^2$ which gives an exponentially small value of the tight-binding wave function near the left electrode.

In contrast, the sub-barrier scattering approach that takes into account the many-electron interactions gives a correct picture of resonant electron transfer. Let us consider the same molecular wire used for the tight-binding exercise above. The parametrization of the scattering amplitude off the individual center is

$$a(\varepsilon) = \frac{1.0}{\varepsilon + 0.037} + (-10 + 350\varepsilon) \tag{34}$$

where the parameters of the pole part in (34) were chosen based on the tight-binding parameters used above (see relationship (30)). The potential part was approximated based on our previous calculations of the scattering amplitudes for different systems. For simplicity, we also neglected the angular dependence of the scattering amplitude in (34).

The numerical solution of the secular determinant (26) gives the energy of the lowest state in the bound spectrum to be in resonance with the Fermi level of the left electrode at applied bias $V = 1$ eV. The scattering vector $\vec{T}$ corresponding to this energy has appreciable component $T_1$ near the left electrode, and the ratio $T_1/T_{30} \approx 2. \times 10^{-5}$ is much larger than that of the tight-binding solution, see Fig. 7. In contrast to single electron potential methods the sub-barrier scattering approach takes into account many-electron interactions and, as a result, the resonant wave function decays much more slowly towards the left electrode. Therefore, resonant tunneling transfer does have an appreciable amplitude and we might expect a new mode of the electron transport that greatly assists the transfer of electrons in the case of relatively long molecules.

The reason for the slower decay of scattering vector components $T_l$ towards the left electrode can be traced back to mutually compensating energy dependences of the scattering operator off the individual center $t(\varepsilon)$ and the Green's function $G_V(R_i, R_{i\pm1}; \varepsilon)$ connecting nearest neighbor centers. This results in a weak energy dependence of the matrix elements of the secular matrix (26). Because spatial dependence is coupled to energy dependence as a result of referencing of local vacuum levels by electrostatic potential as we go along the wire, we obtain a weak spatial dependence of the solution, i.e. scattering vector $\vec{T}$.

Our calculations of resonant electron transitions in an external electric field are based on the fact that neither the external electrostatic potential nor the current produce substantial changes in the electronic density compared to the state of the molecule in the absence of the electric field. This is because the occupation of the charged resonant states contributing to the current is very low. However, recent calculations based on NEGF formalism revealed substantial changes in the charge density and the corresponding electrostatic potential due to an applied electric field and the current passing through the molecule [26]. The states contributing to the current in NEGF theory are the hole states, i.e. occupied states of the neutral molecule that start to participate in transport when an applied bias raises them above the Fermi energy. It is not surprising that these "charged states" (i.e. the states that correspond to the Hamiltonian of the positively charged hole) are very sensitive to the applied electric field and their wave functions are substantially deformed in the electric field (e.g. see the tight-binding wave-function in Fig. 7). However, the occupations of the hole states (on the order of $T_1/T_N$) are very low due to the rapid emptying of the levels that lie in the energy interval $\varepsilon_F - |V| < \varepsilon_s < \varepsilon_F$. Therefore, it is not completely clear why the states with a very small occupation probability are contributing in a substantial way to the charge density and electrostatic potential of the molecule in NEGF calculations of the transport through the molecule.

## 6. Resonant tunneling current

The resonant tunneling current is determined by the total resonant tunneling amplitude

$$A_{tr}^{res}(\varepsilon) = \sum_s A(\varepsilon, \varepsilon_s), \tag{35}$$

which includes contributions from each resonant state $s$ within the energy interval $\varepsilon_F^l - |V| \leq \varepsilon_s \leq \varepsilon_F^l$. For a given $s$ only a narrow energy interval around $\varepsilon_s$, $\varepsilon_s - \gamma_s < \varepsilon < \varepsilon_s + \gamma_s$, contributes to the resonant tunneling transitions because each partial amplitude $A(\varepsilon, \varepsilon_s)$ contains a dominant pole factor $(\varepsilon - \varepsilon_s + i\gamma_s)^{-1}$. Beyond this energy interval only ordinary tunneling takes place, but because of its exponentially small values for sufficiently long molecules, we can neglect its contribution and consider only transitions via resonant energy level $\varepsilon_s$ closest to a given energy $\varepsilon$.

By substituting total resonant tunneling amplitude (35) and (32) in the general expression for the tunneling current (2), the total resonant tunneling current is obtained as a sum of partial currents from each bound state within an energy interval $\varepsilon_F^l - |V| \leq \varepsilon_s \leq \varepsilon_F^l$ and is given by

$$I_{lr}^{res}(V) = \sum_{\varepsilon_F^l - |V| \leq \varepsilon_s \leq \varepsilon_F^l} I_{res}(\varepsilon_s), \tag{36}$$

where

$$I_{res}(\varepsilon_s) = \frac{2\pi^2}{\gamma_s} \widetilde{A}_l^2 \widetilde{A}_r^2 T_1^2(\varepsilon_s) T_N^2(\varepsilon_s) \rho_l(\varepsilon_s) \rho_r(\varepsilon_s). \tag{37}$$

In deriving (36) and (37) we assumed that the width $\gamma_s$ is smaller than the separation between neighboring resonant energy levels $\{\varepsilon_s\}$ which allowed us to replace the Lorentz function by the $\delta$-function: $\left((\varepsilon - \varepsilon_s)^2 + \gamma_s^2\right)^{-1} \to \pi / \gamma_s \delta(\varepsilon - \varepsilon_s)$.

The expression (36) for the resonant tunneling current contains the resonant width $\gamma_s$ that must be determined in a self-consistent manner. The inverse lifetime $\gamma_s$ is determined by the probability of the transition of the electron from state $\varepsilon_s$ to all other states. The Fermi golden rule gives the following expression for this probability $\gamma_s$

$$\gamma_s = 2\pi \left\{ \widetilde{A}_l^2(\varepsilon_s) T_1^2(\varepsilon_s) + \widetilde{A}_r^2(\varepsilon_s) T_N^2(\varepsilon_s) \right\} \tag{38}$$

where the first and second terms in (38) give the probabilities of transitions from state $s$ to the left and to the right electrodes respectively. We have already learned that the left component of the scattering vector $T_1$ is much smaller than the right component $T_N$, therefore the first term can be dropped in (38). Then, we obtain the final expression for the partial resonant tunneling current due to resonant state $\varepsilon_s$

$$I_{res}(\varepsilon_s) = \pi \tilde{A}_l^2(\varepsilon_s) T_1^2(\varepsilon_s) \rho_l(\varepsilon_s). \tag{39}$$

The apparent asymmetry of the expression (39) is due to the presence of a rate limiting step in the resonant transfer which is a transfer of the electron from one of the states of the left electrode to the resonant bound state $s$. Once the electron reaches the virtual bound state $s$, then this level is quickly emptied by a fast transfer to one of the states of the right electrode.

The current-voltage curve in resonant tunneling regime would exhibit a step-like structure as the bias is increased because more and more resonant states from the bound state spectrum are included. However, due to temperature effects, the step-like structure is smeared out as a result of the fluctuations in the resonant energy levels around zero-temperature values $\varepsilon_s^0$. This effect is present even at low temperatures because positions of the energy levels depend exponentially on the distance between the nearest-neighbor centers $d$, see (26) and the effect of small molecular vibrations to be greatly amplified.

We calculated the resonant I-V curve for the same model system that we considered in section 4 using parametrization (34) for the scattering amplitude $a(\varepsilon)$. The calculated I-V curve shown in Fig. 8 has a distinct threshold $V_{th} \approx 1\,\text{eV}$, the currents being on the order of nA which is in quantitative agreement with experiment on single strand DNA [13]. In another experiment [12] bundles of DNA were used in measurements and no threshold was observed. Instead, I-V curves showed Ohmic behavior, i.e. a linear increase of the current with the applied bias. In order to take into account the conditions of experiment, we assumed that the weak inter-strand interaction in the bundle will only slightly modify the scattering operator. Therefore, we increased a linear slope in the potential part of the scattering amplitude by ~ 10%

$$a(\varepsilon) = \frac{1}{\varepsilon + 0.037} + (-10 + 400\varepsilon). \tag{40}$$

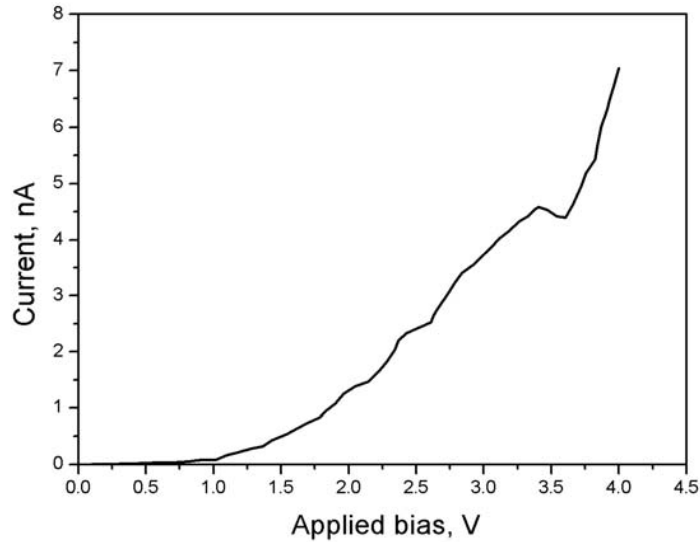

Fig. 8. Resonant I-V curve for a model of single strand of DNA, that shows clear threshold behavior, was calculated using the parameters corresponding to experiment [12].

As in experiment, we were specifically interested in the range of small applied biases $0 \leq V \leq 0.1\,\text{V}$. A calculated I-V curve for this case is shown in Fig. 9. In contrast to the previous case, we did not observe a threshold in the I-V curve. Moreover, the conductance $dI/dV$ is constant, i.e. the I-V curve is linear and the regime is indeed Ohmic. Also, the values of the current at such small bias are much higher as compared to the case of single strand DNA.

The resonant mode of transport is usually characterized by a weak length dependence of the tunneling current on the distance between the left and right electrodes. Therefore, we investigated the length dependence of the conductance. The I-V curve was calculated for a molecule consisting of $N = 60$ and compared to the case $N = 30$. We did not observe the length dependence of the conductance, i.e. the I-V curves are very similar.

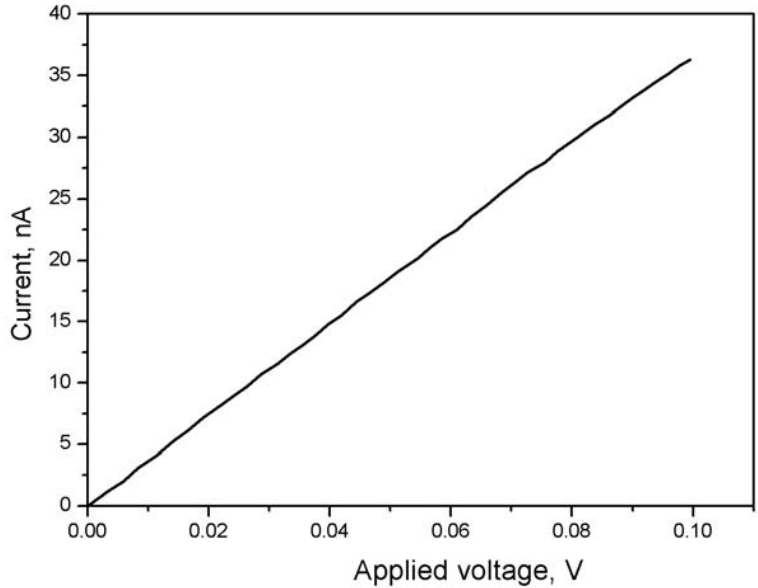

Fig. 9. Ohmic I-V curve that was calculated for a model system that mimic the transport in DNA bundles.

We explained the substantial conductivity of relatively long DNA molecules with the dominant contribution of the resonant electron transfer to electron transport. Recently, a new mechanism of charge transport in DNA molecules due to hopping of the holes has been proposed by Jortner and co-workers [27,28]. The variable-range hoping is based on the assumption that there are electron (or hole) traps in the medium due to electronic defects. In the case of DNA the electron energies of these defect levels with respect to vacuum are comparable to the ionization potential of DNA, i.e. they lie several eV below the Fermi energies of the electrodes. Therefore, this mechanism is not operational for the case of electron transport in metal-molecule-metal systems. An additional experimental confirmation of this statement is a weak temperature dependence of the DAN conductance [29]. If variable-range hopping were important it would exhibit a strong temperature dependence due to its activation nature.

## 7. Conclusions

In this paper we presented a new approach for investigating electron transport through organic molecules. This theory is drastically different from the standard, one-electron potential description of the electron structure widely used to model electron transport through single molecules. We found that the many-electron effects play an important role in electron transport and in order to address them, we have developed a theory of sub-barrier scattering that includes exchange interactions naturally within its remit. Our approach predicted two important mechanisms of electron transport: ordinary tunneling and resonant tunneling. In the first case, sub-barrier-scattering theory predicts a substantial amplification of the tunneling current by the molecule (bridge) compared to vacuum tunneling, the amplified tunneling exponents being in good agreement with experiment. The physics of resonant tunneling is determined by the bound energy spectrum of the tunneling electron. Based on the position of the lowest level of the band in respect to the Fermi energy of one of the electrodes, we predicted threshold and Ohmic modes of transport. Although we illustrated the features of transport mechanisms by performing model calculations, we are confident that several aspects of electron transport are fundamental phenomena that will also be present in more elaborate calculations that we plan to do in the future.

## Acknowledgments

M.A.K. and V.S.P. thank the Russian Foundation for Basic Research for financial support under grant 05-03-32102. I.I.O. thanks the National Science Foundation for financial support under grant CCF-0432121.